\documentclass[showpacs,preprint,prb,aps,superscriptaddress]{revtex4}
\usepackage{graphicx}
\usepackage[normalem]{ulem}

\begin{document}

\title{Raman spectra in iron-based quaternary CeO$_{1-x}$F$_x$FeAs and LaO$_{1-x}$F$_x$FeAs}
\author{S. C. Zhao}
\affiliation{Department of Physics,
Renmin University of China, Beijing 100872, P. R. China}
\author{D. Hou}
\affiliation{Department of Physics,
Shandong University, Jinan 250100, P. R. China}
\affiliation{Department of Physics,
Renmin University of China, Beijing 100872, P. R. China}
\author{Y. Wu}
\affiliation{National Laboratory of Solid State
Microstructures, Department of Physics, Nanjing University,
Nanjing 210093, P. R. China}
\author{T. L. Xia}
\author{A. M. Zhang}
\affiliation{Department of Physics,
Renmin University of China, Beijing 100872, P. R. China}
\author{G. F. Chen}
\author{J. L. Luo}
\author{N. L. Wang}
\affiliation{Beijing National Laboratory for Condensed Matter
Physics, Institute of Physics, Chinese Academy of Sciences,
Beijing 100080, P. R. China}
\author{J. H. Wei}
\author{Z. Y. Lu}
\affiliation{Department of Physics,
Renmin University of China, Beijing 100872, P. R. China}
\author{Q. M. Zhang}
\email{qmzhang@ruc.edu.cn}
\affiliation{Department of Physics,
Renmin University of China, Beijing 100872, P. R. China}
\affiliation{National Laboratory of Solid State Microstructures,
Department of Physics, Nanjing University, Nanjing 210093, P. R.
China}

\date{\today}

\begin{abstract}
Raman spectra have been measured on iron-based quaternary
CeO$_{1-x}$F$_x$FeAs and LaO$_{1-x}$F$_x$FeAs with varying fluorine doping
at room temperatures. A group analysis has been made to clarify the
optical modes. Based on  the first principle calculations, the observed
phonon modes can be assigned accordingly. In LaO$_{1-x}$F$_x$FeAs, the
E$_g$ and A$_{1g}$ modes related to the vibrations of La, are suppressed
with increasing F doping. However F doping only has a small effect
on the E$_g$ and A$_{1g}$ modes of  Fe  and As. The Raman modes of La
and As are absent  in rare-earth substituted CeO$_{1-x}$F$_x$FeAs, and
the E$_g$ mode of oxygen, corresponding to the in-plane vibration of
oxygen, moves to around 450 cm$^{-1}$ and shows a very sharp peak.
Electronic scattering background is low and electron-phonon coupling
is not evident for the observed phonon modes. Three features are found above
500 cm$^{-1}$, which may be associated with multi-phonon process. Nevertheless
it is also possible that they are related to magnetic fluctuations or interband
transitions of d orbitals considering their energies.
\end{abstract}


\maketitle

\section{Introduction}
The discovery of  iron-based superconductors has stimulated many
interests in searching higher-T$_c$ superconductors without
copper-oxide planes. Soon after the first LaO$_{1-x}$F$_x$FeAs was
reported with T$_c$=26 K\cite{kamihara}, rare-earth substitution
compounds SmO$_{1-x}$F$_x$FeAs and CeO$_{1-x}$F$_x$FeAs were
synthesized and their T$_c$'s can go up to above 40 K \cite{X. H.
Chen,G. F. Chen}. For LaO$_{1-x}$F$_x$FeAs, T$_c$ can be raised to
above 40 K under high pressure\cite{Takahashi}. A higher T$_c$ of 52
K in Fe-based series superconductors was reported in
PrO$_{1-x}$F$_x$FeAs which was synthesized under high
pressure\cite{Z. A. Ren}. The measurements of Hall coefficients
indicates that the carriers in the superconductors are
electron-like. On the other hand, hole-doped (La,
Sr)O$_{1-x}$F$_x$FeAs was synthesized successfully with T$_c$=26
K\cite{H. H. Wen}. Transition-metal substituted LaO$_{1-x}$F$_x$NiAs
was also reported with T$_c$=4 K and an extremely sharp
superconducting transition\cite{Z. Li}. Most recently, a very
exciting result shows that a maximum T$_c$ of 55K can be obtained
even without F doping in the Fe-based superconductors\cite{Z. A.
Ren3} . Up to now the record of the highest T$_c$ 56.5 K in
iron-based superconductors was archived in Gd$_{1-x}$Th$_x$OFeAs
without F doping\cite{C. Wang}. It is believed that electrons are
transferred into FeAs conducting layers effectively by Th$^{4+}$
substitution of Gd$^{3+}$, just like F doping. This means that
carrier concentration and superconducting transition temperature can
be controlled and tuned by oxygen content and rare-earth
substitution, which is much similar to the case of cuprate
superconductors. A very high upper critical field over 100T was
estimated by resistance measurements under high magnetic field, even
exceeding that of cuprate superconductors\cite{Hunte}.

The early band structure calculations suggested that the pure
LaOFeAs compound is a nonmagnetic metal but with strong
ferromagnetic or antiferromagnetic (AFM) instability \cite{singh}.
Later, it was found that the antiferromagnetically ordered
state\cite{C. Cao,F. J. Ma} has a lower energy than the nonmagnetic
one, due to the Fermi surface nesting\cite{F. J. Ma}. It was then
\cite{J. Dong} predicted that the AFM state should form a striped
structure by breaking the rotational symmetry, which was indeed
observed by the neutron scattering experiment\cite{Clarina,McGuire}.
The further theoretical studies show that the superexchange
antiferromagnetic interaction between the next nearest neighbor
Fe-Fe bridged by As is responsible for the stripe-ordered Fe
moments\cite{yild,ma2}. Some microscopic models have been
constructed to understand the superconductivity in the compounds
with iron-arsenite plane instead of copper-oxygen plane. There is no
consensus on some important issues such as pairing mechanism and
symmetry yet. More experimental efforts and further theoretical
considerations are needed to reveal the properties of ground state
and magnetic excitations.

 For the parent compounds
without F doping,  a subtle  structural change and a successive
magnetic transition were revealed by means of transport
measurements, specific heat and neutron scattering etc. It is still
a key issue to understand the connection between the structural
change and magnetic transition. Furthermore, exploring the driving
force for pairing requires a basic knowledge on some important
interactions such as  electron-phonon coupling. So it is important
and necessary to learn more detailed information on lattice
vibrations in the compounds. Raman scattering is known as a unique
technique in studying optical phonon modes. Besides, if allowed by
selection rules, it is possible to obtain additional information
about magnetic, electronic and other collective excitations. In
fact, a nice Raman scattering study on iron-based materials has been
done on ab plane of single crystal grain by microscopic Raman
method\cite{Hadjiev}. However the spectra only covered 100 to 400
cm$^{-1}$ and in-plane vibrations were not resolved due to the limit
of ab plane.

In this paper, Raman scattering measurements at room temperature
from 30 to 2000 cm$^{-1}$, have been performed on two kinds of
iron-based materials CeO$_{1-x}$F$_x$FeAs with  x=0 (T$_c$=0 K) and
0.16(T$_c$=41 K),  and LaO$_{1-x}$F$_x$FeAs with  x=0 (T$_c$=0 K),
0.04(T$_c$=17 K) and x=0.08 (T$_c$=26 K), respectively. Six Raman
phonon modes were observed below 500 cm$^{-1}$. Based on the
structural data obtained by neutron scattering and X-ray
diffraction, a group analysis was made to classify the optical
modes. Then first-principle calculations were carried out to
calculate the optical modes at $\Gamma$ point. By comparison with
the calculations and group analysis,  the observed phonon modes were
assigned accordingly. The changes of some modes with F doping and
rare-earth substitution were discussed. And electron-phonon coupling
was found to be small. Three weak features were observed above  500
cm$^{-1}$, which may be associated with multi-phonon process,
magnetic fluctuations or interband transitions.

\section{experimental details}

Polycrystal samples of CeO$_{1-x}$F$_x$FeAs and LaO$_{1-x}$F$_x$FeAs
were synthesized by solid state reaction method. As precursor
materials, CeAs or LaAs were presynthesized by reacting Ce or La
chips and As pieces in an evacuated quartz tube. Then FeAs was
obtained with a similar process. Using CeAs/LaAs, Fe,
CeO$_2$/La$_2$O$_3$, CeF$_3$/LaF$_3$ and FeAs as starting materials,
the raw materials with stoichiometric ratio were mixed thoroughly
and pressed into pellets. The pellets were warped with Ta foil and
sealed in an evacuated quartz tube. Almost pure phase polycrystal
samples were obtained after annealing. The detailed procedure for
preparing the samples can be found elsewhere\cite{G. F. Chen}. The
temperature dependence of susceptibility and resistivity of the
polycrystal samples used in the present Raman study is shown in
Fig.~1. The transition temperature width shows the high quality of
the samples. For CeO$_{1-x}$F$_x$FeAs, many shining single crystal
grains can be seen even without a microscope. After polishing, a
flat alloy-like surface can be obtained. Unfortunately Raman signal
is too weak to be detected because most intensities of excitation
light were reflected back by the alloy-like surface. In order to
increase the intensities of incident light effectively and to
resolve more phonon modes beyond ab plane,  the present Raman
measurements were done with ground fine powder.

The Raman measurements were performed with a triple-grating
monochromator (Jobin Yvon T64000), which works with a microscopic
Raman configuration. A 50$\times$ objective microscopic lens with a
working distance of 10.6 mm, is used to focus the incident light on
sample and collect the scattered light from sample. The detector is
a back-illuminated CCD cooled by liquid nitrogen. An solid-state
laser (Laser Quantum Torus 532) with high-stability and very narrow
width of laser line, is used with an excitation wavelength of 532
nm. The laser beam of 3 mW was focused into a spot of less than 10
microns in diameter on sample surface.

\section{Assignment of Raman Phonons}

The refined structure parameters of LaOFeAs have been obtained by
neutron scattering and X-ray diffraction. It has a tetragonal
ZrCuSiAs-type structure with space group P4/nmm (No. 129, origin
choice 2) and point group D$_{4h}$. Atoms La, O/F, Fe and As occupy
\begin{center}
Table I: Classifications of optical modes of LaOFeAs

\begin{tabular}{|c|c|c|c|}\hline\hline
Atom & Wyckoff position & Raman modes & IR modes \\\hline
La & 2c & A$_{1g}$+E$_g$ & A$_{2u}$+E$_u$ \\
O/F & 2a & B$_{1g}$+E$_g$ & A$_{2u}$+E$_u$ \\
Fe & 2b & B$_{1g}$+E$_g$ & A$_{2u}$+E$_u$ \\
As & 2c & A$_{1g}$+E$_g$ & A$_{2u}$+E$_u$ \\\hline\hline
\end{tabular}
\end{center}
Wyckoff positions 2c, 2a, 2b and 2c, respectively. Symmetry
analysis shows that there are eight Raman-active modes and six
infrared(IR)-active modes, as classified in Table I.

\

Table II: Assignment of optical phonons in comparison with the
first-principle  calculations. The cited data of IR phonons comes
from ref. 14. Just the dominant atoms involved in a specific mode
are listed here. The last column describes the approximate
direction and phase of vibration, where [100], [010], [001] are
parallel to the a, b, c crystal axis, respectively. See FIG. 3 for
the crystal coordinate system.

\ \ \ \begin{tabular}{|p{2.5cm}|p{2cm}|c|c|c|l|}\hline\hline
Experimental (cm$^{-1}$) & Calculated (cm$^{-1}$) & Symmetry &
Active & Atom & Vibration \\\hline
                 & 60.95657  & E$_{u}$  & IR    & La, As, Fe  & [100] or [010]\\
\ \ \ \ \ \ \ 97 & 80.73106  & A$_{2u}$ & IR    & La, As, Fe  & [001]\\
\ \ \ \ \ \ \ 96 & 111.02250 & E$_{g}$  & Raman & La        & [100] or [010], out-of-phase\\
 \ \ \ \ \ \ 137 & 138.06264 & E$_{g}$  & Raman & As, Fe    & [110] or [1-10], out-of-phase \\
 \ \ \ \ \ \ 161 & 180.56708 & A$_{1g}$ & Raman & La        & [001], out-of-phase \\
                 & 203.74154 & A$_{1g}$ & Raman & As        & [001], out-of-phase \\
 \ \ \ \ \ \ 214 & 218.21930 & B$_{1g}$ & Raman & Fe        & [001], out-of-phase\\
 \ \ \ \ \ \ 248 & 254.15852 & A$_{2u}$ & IR    & Fe, As    & [001] \\
 \ \ \ \ \ \ 266 & 268.08770 & E$_{u}$  & IR    & Fe, As    & [100] or [010] \\
 \ \ \ \ \ \ 278 & 279.24677 & E$_{g}$  & Raman & Fe        & [110] or [1-10], out-of-phase \\
                 & 280.21650 & E$_{u}$  & IR    & O         & [100] or [010], in-phase \\
                 & 282.44946 & B$_{1g}$ & Raman & O         & [001], out-of-phase \\
 \ \ \ \ \ \ 338 & 388.45420 & A$_{2u}$ & IR    & O         & [001], in-phase\\
 \ \ \ \ \ \ 423 & 420.80567 & E$_{g}$  & Raman & O         & [100] or[010], out-of-phase\\
\hline\hline
\end{tabular}

\

To assign the phonon modes at $\Gamma$-point, the phonon frequencies
of non-magnetic LaOFeAs crystal have been calculated in the
framework of the density perturbation functional theory (DFPT) using
plane-wave pseudopotentials \cite{QE} with the generalized gradient
approximation (GGA)of Perdew-Burke-Ernzerh (PBE)\cite{PBE} for the
exchange-correlation potentials. Firstly, a self-consistent
calculation was accomplished using the experimental lattice
parameters and the energy minimized internal atomic
positions\cite{F. J. Ma} , during which the following parameters
were adopted: the 45 and 360 Ryd cutoffs of kinetic energy for
wave-function and charge-density, respectively; the
24$\times$24$\times$12 uniform k-space for the k-points integration;
and the gaussian smearing parameter of 0.002 Ryd. Then, by
diagonalizing the dynamical matrix generated from converged
potentials, we obtained the eigenvalues, namely the squared phonon
frequencies, and the eigenvectors to derive the displacements of
each atom in a specific vibration mode.

Eight frequency-distinguishable Raman-active phonon modes were
deduced at $\Gamma$-point, consisting of four two-fold degenerated
E$_g$, two non-degenerated A$_{1g}$ and two non-degenerated B$_{1g}$
modes, which consists with the group symmetry analysis. Comparing
them to above Raman spectra showed a fairly good agreement, see Fig.
2 and Tab. II. All the Raman active vibration modes are
characterized in Fig. 3, where the arrows indicate the vibration
directions of the corresponding atoms, with their lengths
representing the relative vibration amplitude compared with those of
other atoms in the same mode.

\section{Discussions}
In Fig. 2, at first glance LaOFeAs system  has a similar phonon
spectrum for different F doping. There is no obvious frequency shift
or change of peak shape for the three samples. However there still
exist some small changes with varying F doping. Both E$_g$ mode at
96 cm$^{-1}$ and A$_{1g}$ mode at 161  cm$^{-1}$ of La are
suppressed gradually with increasing F doping. It can be naturally
understood as the effect of F dopants entering LaO layer. The
sensitivity of La phonon modes to F doping, may be considered as an
alternative way to characterize the concentration of F doping.

F doping has little effect on the Fe and As-dominated phonon modes,
such as B$_{1g}$ mode at 214 cm$^{-1}$ and E$_g$ mode at 278
cm$^{-1}$ of Fe, and E$_g$ mode at 137 cm$^{-1}$ of As. The fact can
be easily explained because F dopants are far away from FeAs layer
and thus has little effect on Fe and As-related vibrations.   These
three modes have a special significance in exploring the mechanism
of magnetic phase transition near 140 K revealed by some
experiments. As suggested by neutron scattering and other
measurements, there also exists a structural changes just above the
temperatures of magnetic phase transition. For iron-based
superconductors, it is still a key issue to make clear whether there
is a connection between the magnetic transition and structural
change or not. The modes related to Fe and As would play an
important role in answering the above question in further Raman
scattering measurements at various temperatures and with applied
magnetic field.

A relatively broad E$_g$ mode of oxygen was observed near 423
cm$^{-1}$ for LaOFeAs. We will come back to this point in
combination with the results of CeOFeAs in the following.

A striking change in the Raman spectra of CeOFeAs is that the E$_g$
phonon of oxygen becomes very strong and moves to 450 cm$^{-1}$. The
reason for the sharper oxygen E$_g$ mode can be originated from a
better sample quality of CeOFeAs as described above. Interestingly,
Infrared measurements also observed a mode with quite high
intensities near 430 cm$^{-1}$ in both LaOFeAs and CeOFeAs, which
moves to higher frequencies with decreasing temperatures, showing a
typical phonon behavior\cite{G. F. Chen, J. Dong}. However, the
first principle calculations show that there should be no IR-active
mode above 400 cm$^{-1}$. It is speculated that an inversion
symmetry breaking can cause an intensity leakage of oxygen E$_g$
mode into IR channel. If it is true, that means that a subtle
structural distortion could occur even at room temperatures.

The other change is that the  modes below  200 cm$^{-1}$ contributed
by La and As, can not be observed. This may be caused by the
distortion of LaO layer due to the substitution of La by smaller Ce
ions. The Raman results based on single crystal are needed to
confirm this point.

Besides the above first-order Raman phonons, there exist some common
features above 500 cm$^{-1}$ in both LaOFeAs and CeOFeAs, which are
located at 590 cm$^{-1}$, 846 cm$^{-1}$and 1300 cm$^{-1}$.
Generally, these features originate from multi-phonon processes. For
instance, the feature at 846 cm $^{-1}$ can be contributed by two
E$_g$ phonons of oxygen simply considering its frequency. On the
other hand, the energies of features correspond to 73, 105 and 161
meV, respectively. They are close to the interband differences of d
orbitals\cite{Haule}, and also the ferromagnetic/antiferromagnetic exchange
energies according to electronic structure calculations\cite{F. J.
Ma,ma2}. At present it can not be ruled out that some of the
features are associated with the interband transitions or magnetic
fluctuations. Further Raman experiments under applied magnetic field
could be helpful to answer this question.

Although polycrystal samples were used in the present measurements,
it can be seen that electronic scattering background is low. And for
the observed modes, especially those of Fe and As, it is not easy to
distinguish evident features of electron-phonon coupling. The
measurements on single crystal are necessary in the further study on
electronic Raman scattering.

\section{conclusions}

In summary, Raman measurements of newly discovered iron-based
superconductors have been performed at room temperatures. The
observed phonon modes are assigned in combination with symmetry
analysis and first-principle calculations. The phonon modes are
discussed in detail. The assignment will provide a basis for further
study on structural and electronic properties and exploring the
superconductivity in the high-T$_c$ materials. Besides the
first-order Raman phonons, more features found at higher frequencies
may be related to magnetic excitations or interband transitions.
\section{acknowledgments}

The work was supported by the MOST of China (973 project
No.:2006CB601002d \& 2006CB9213001) and NSFC Grant No. 10574064 \&
20673133.

\clearpage

\begin{figure}
\includegraphics[width=12cm]{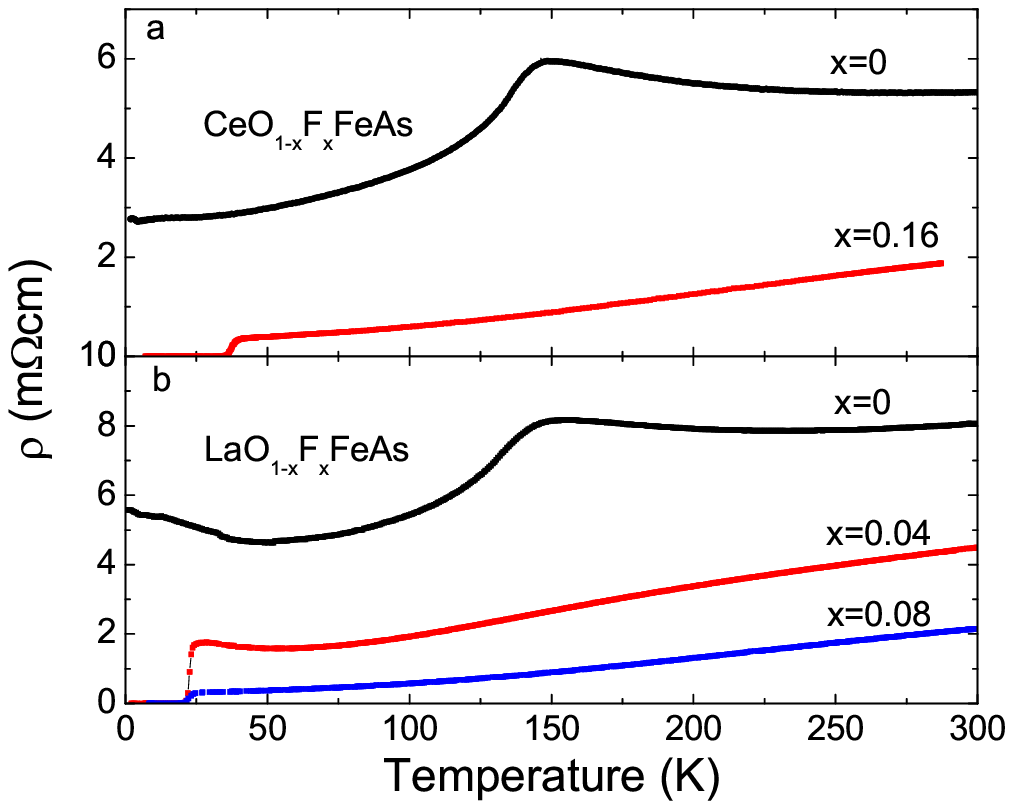}
\caption{(color online) Temperature dependence of resistivity for
(a)CeO$_{1-x}$F$_x$FeAs and (b)LaO$_{1-x}$F$_x$FeAs used in the
present Raman measurements. The sharp superconducting transitions
indicate the high-quality of samples. Especially for the
superconducting CeO$_{1-x}$F$_x$FeAs, it shows a quite small
residual resistivity by extrapolation.} \label{fig1}
\end{figure}

\clearpage

\begin{figure}[tb]
\includegraphics[width=12cm]{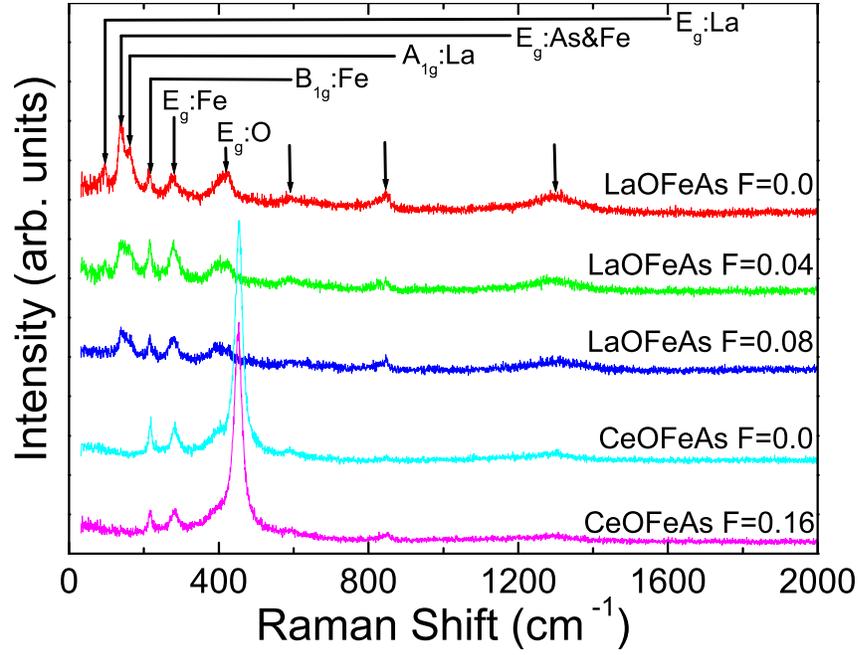}
\caption{(color online) Raman spectra of CeO$_{1-x}$F$_x$FeAs and
LaO$_{1-x}$F$_x$FeAs at room temperatures. } \label{fig2}
\end{figure}

\clearpage

\begin{figure}[tb]
\includegraphics[width=4.0cm,angle=0]{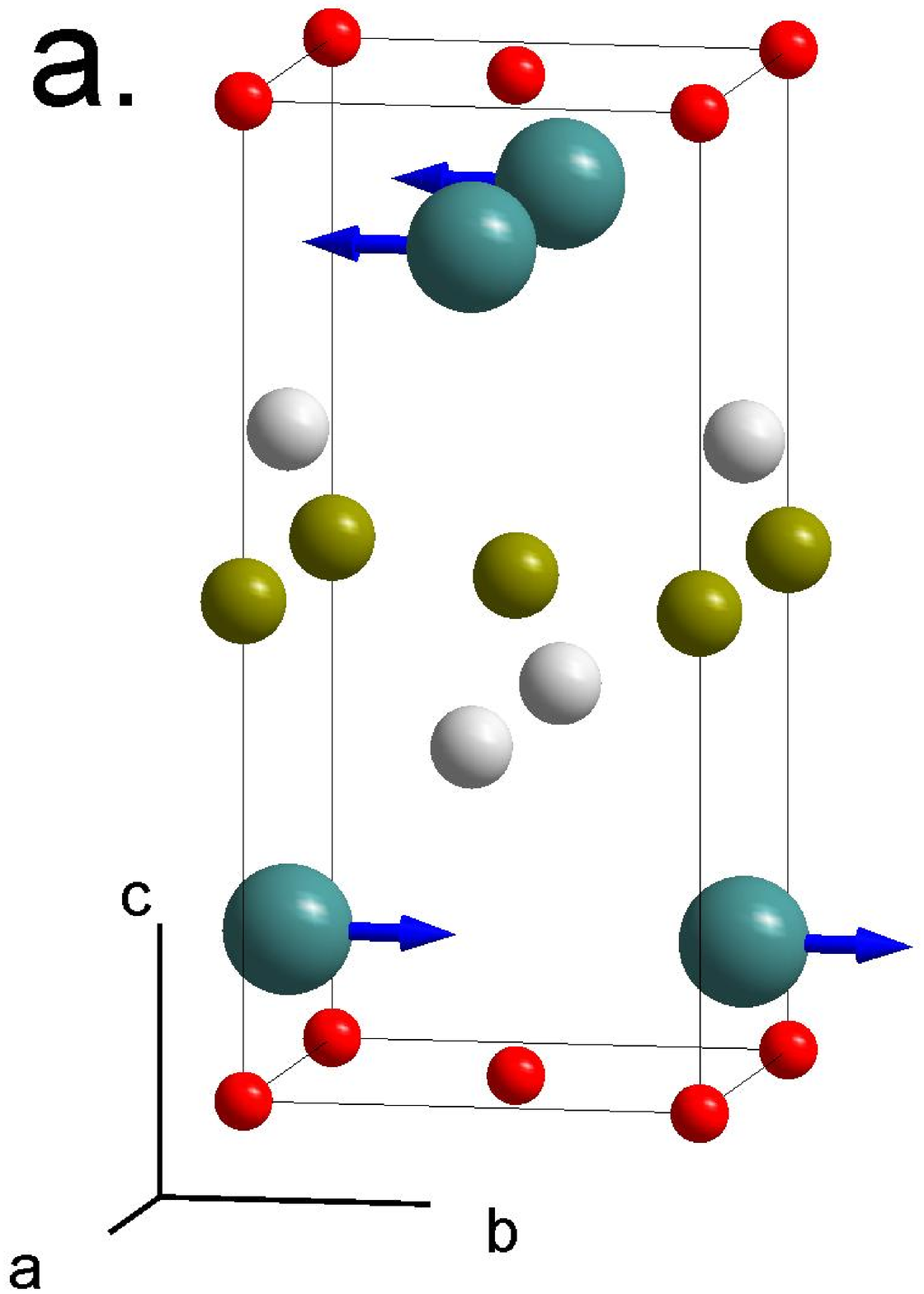}
\includegraphics[width=4.0cm,angle=0]{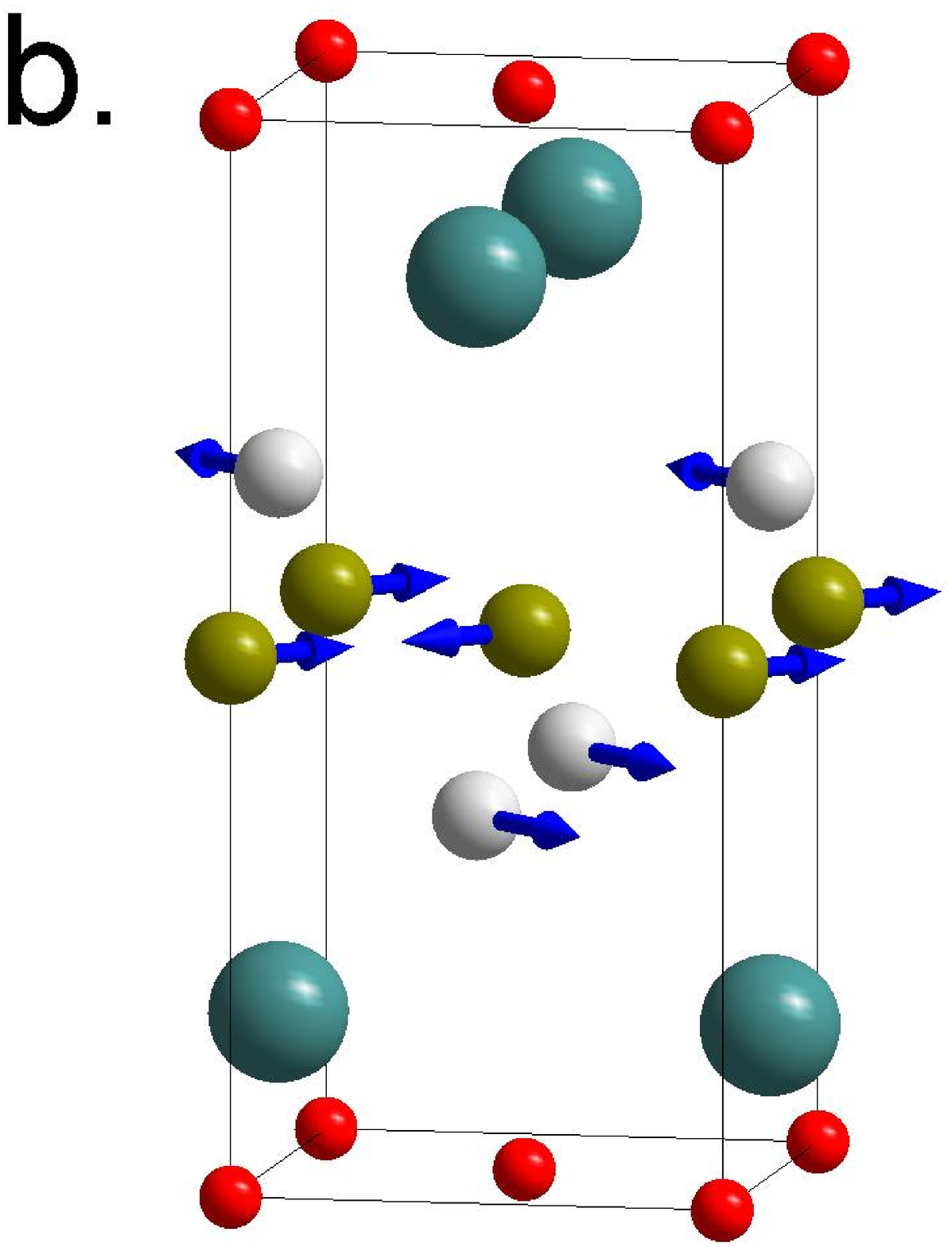}
\includegraphics[width=4.0cm,angle=0]{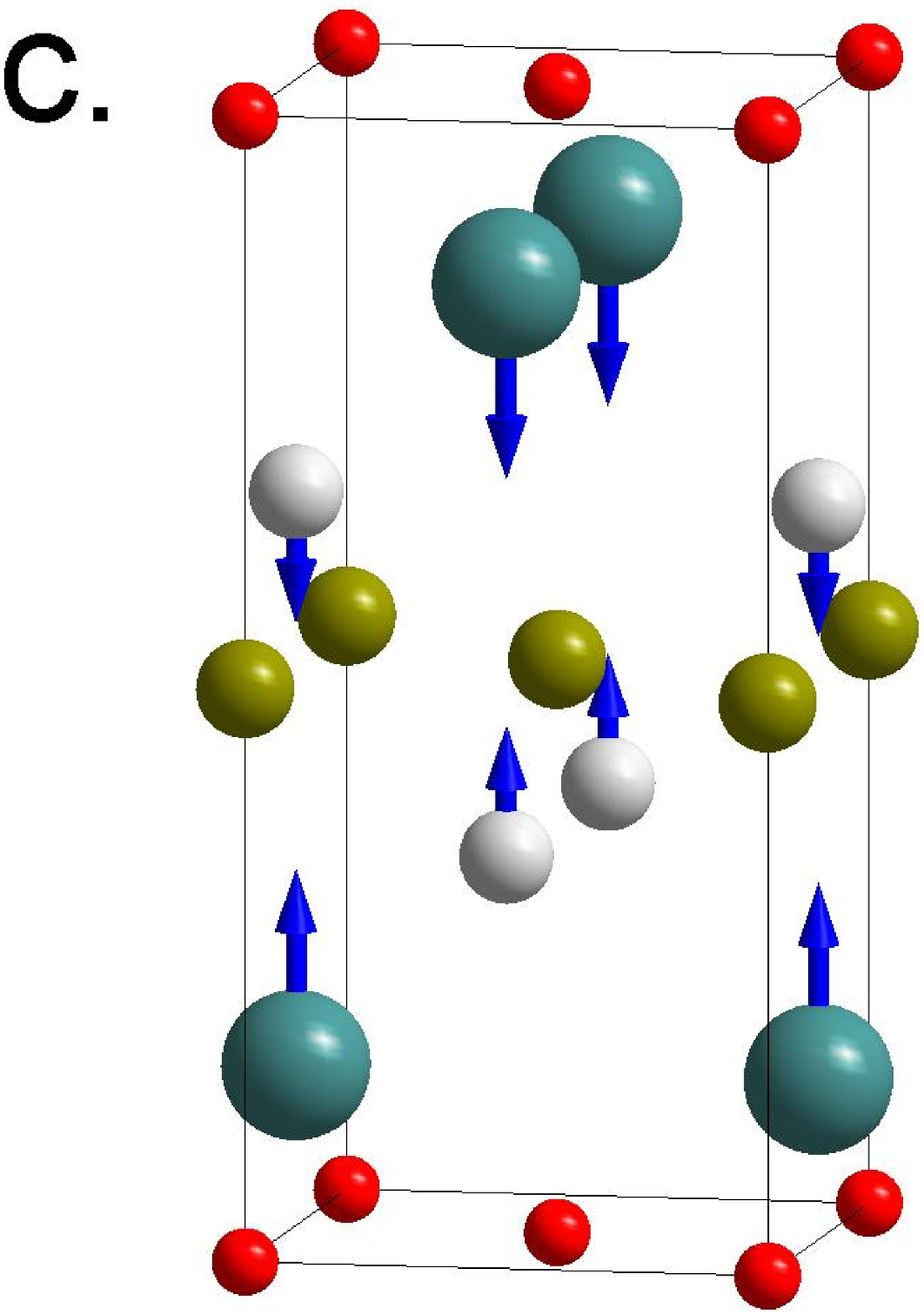}
\includegraphics[width=4.0cm,angle=0]{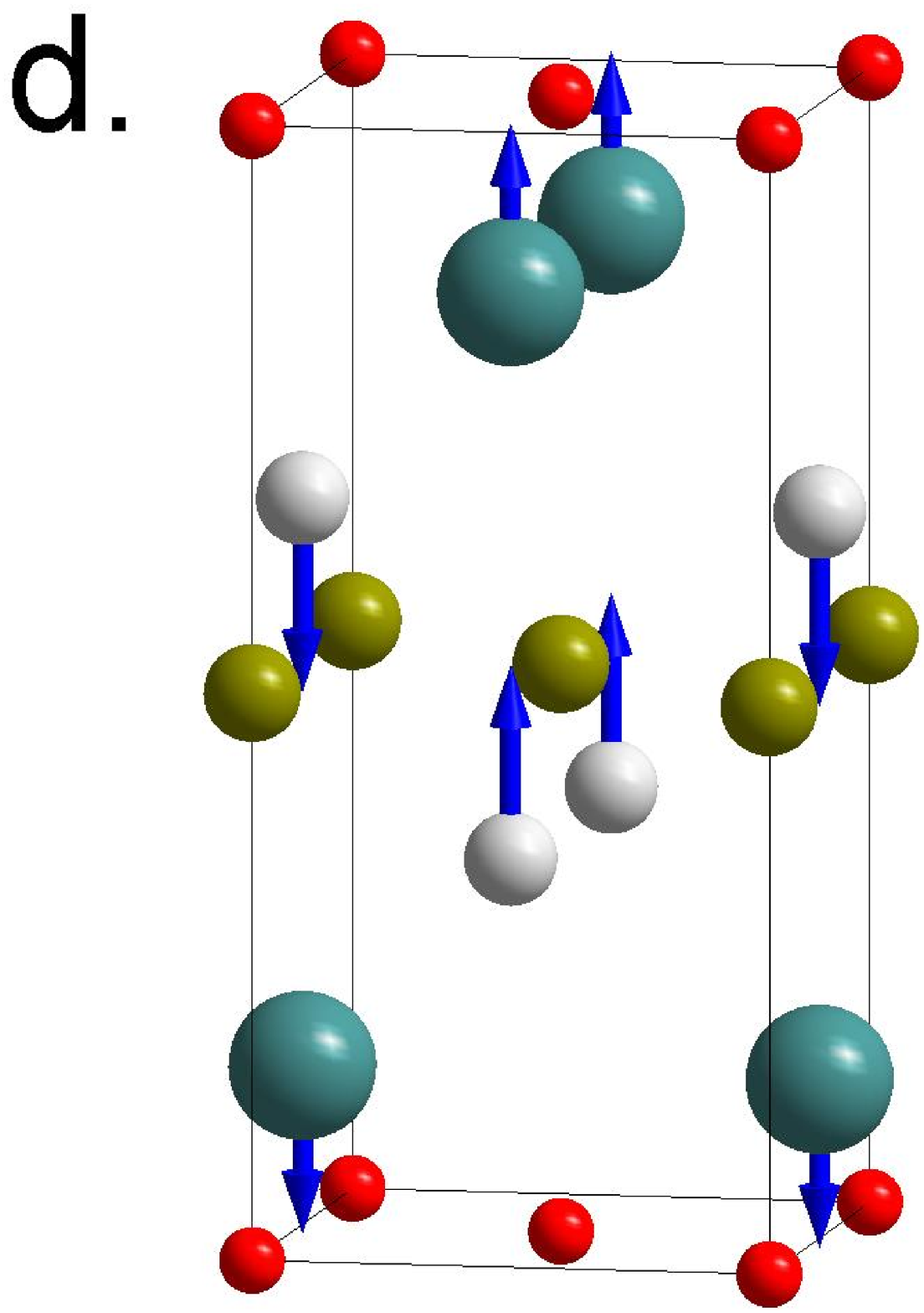}

\includegraphics[width=4.0cm,angle=0]{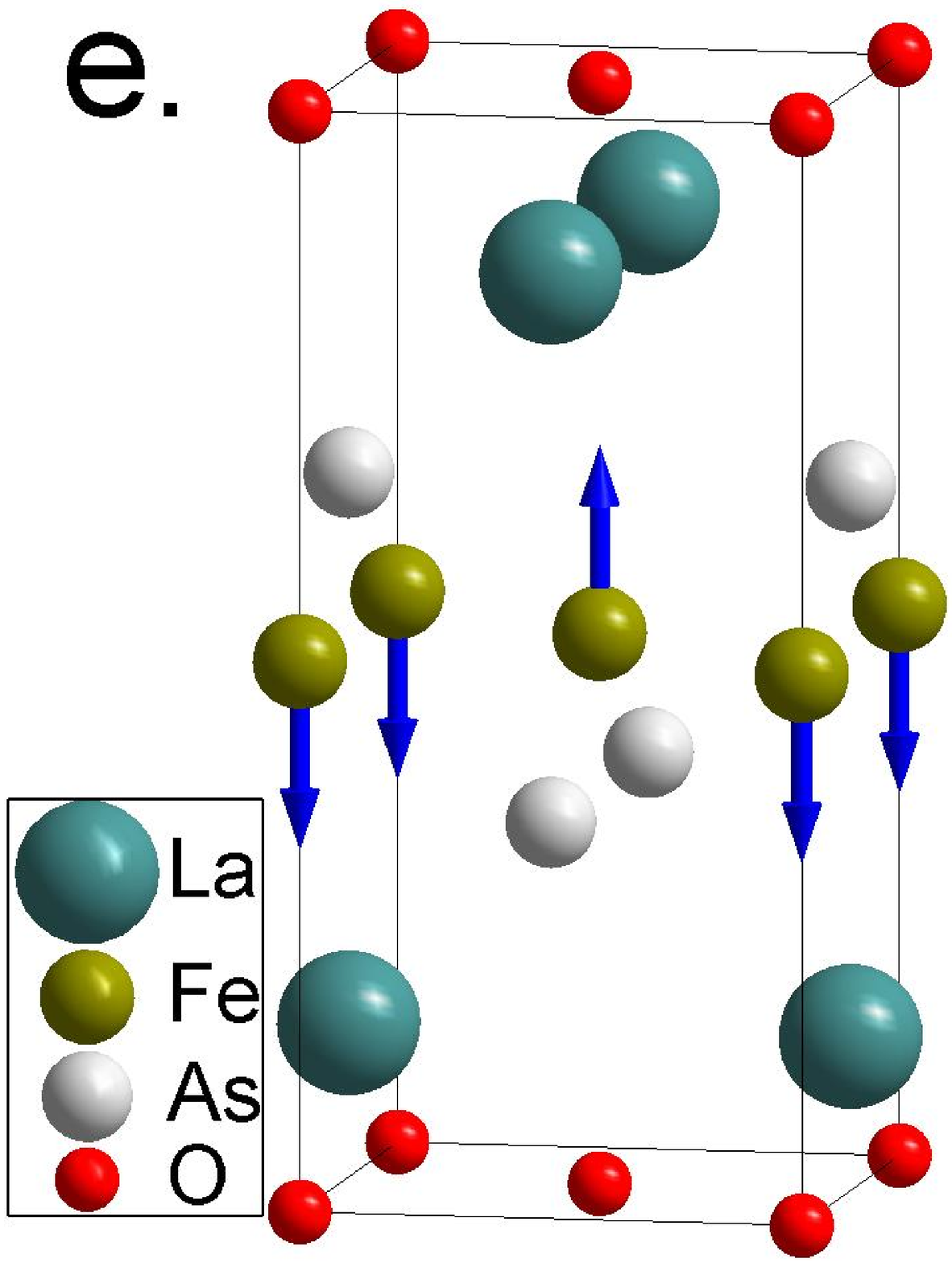}
\includegraphics[width=4.0cm,angle=0]{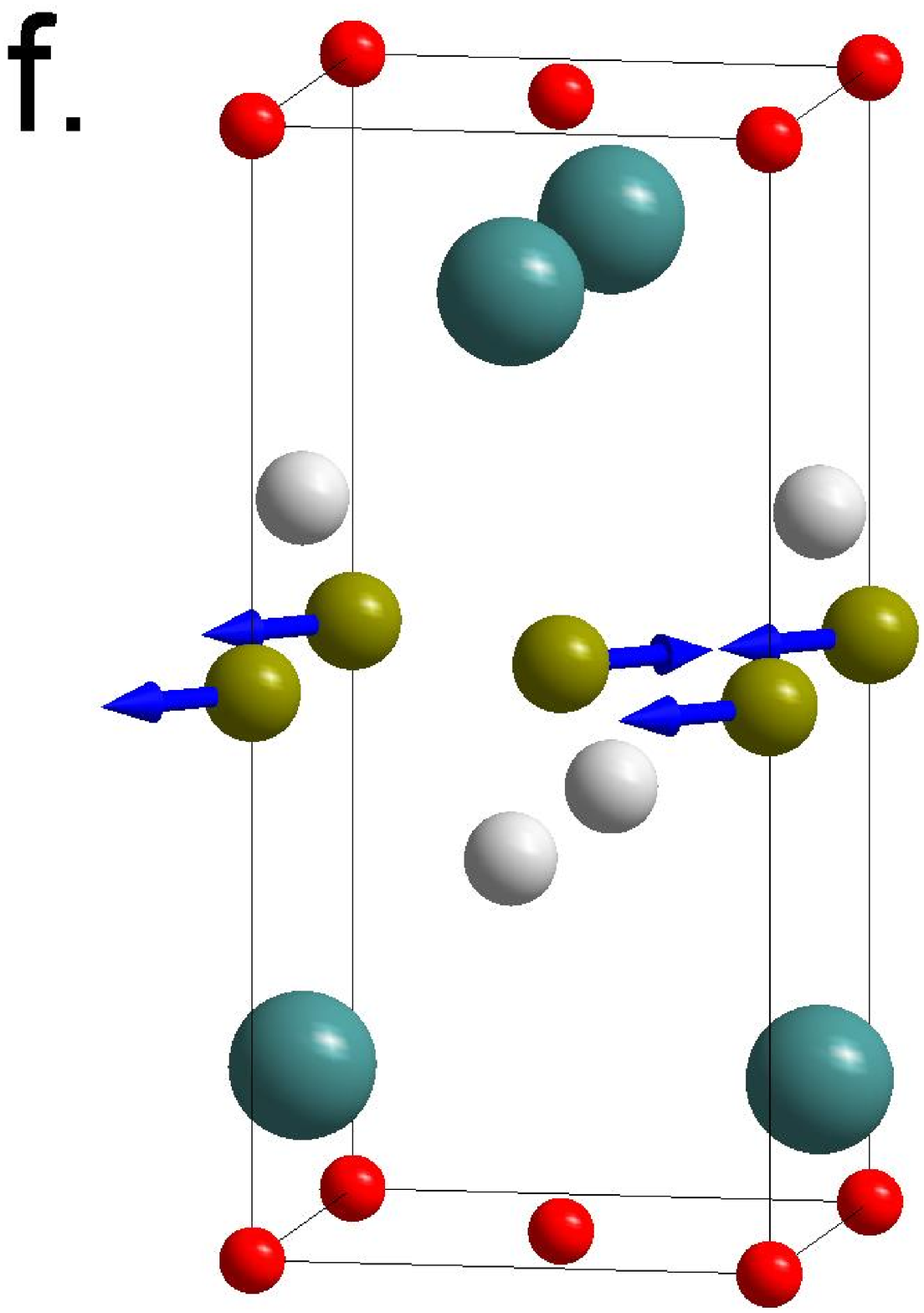}
\includegraphics[width=4.0cm,angle=0]{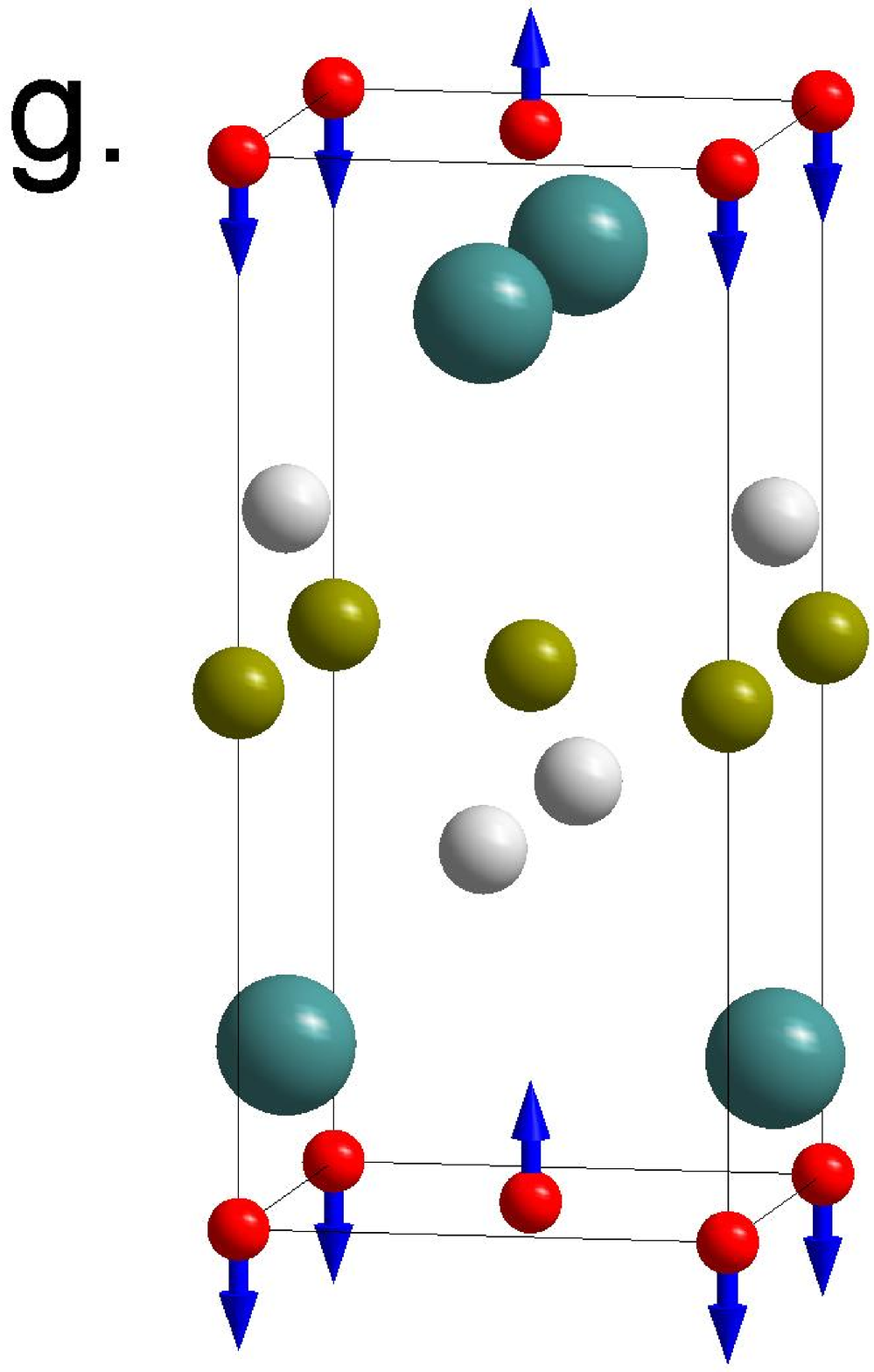}
\includegraphics[width=4.0cm,angle=0]{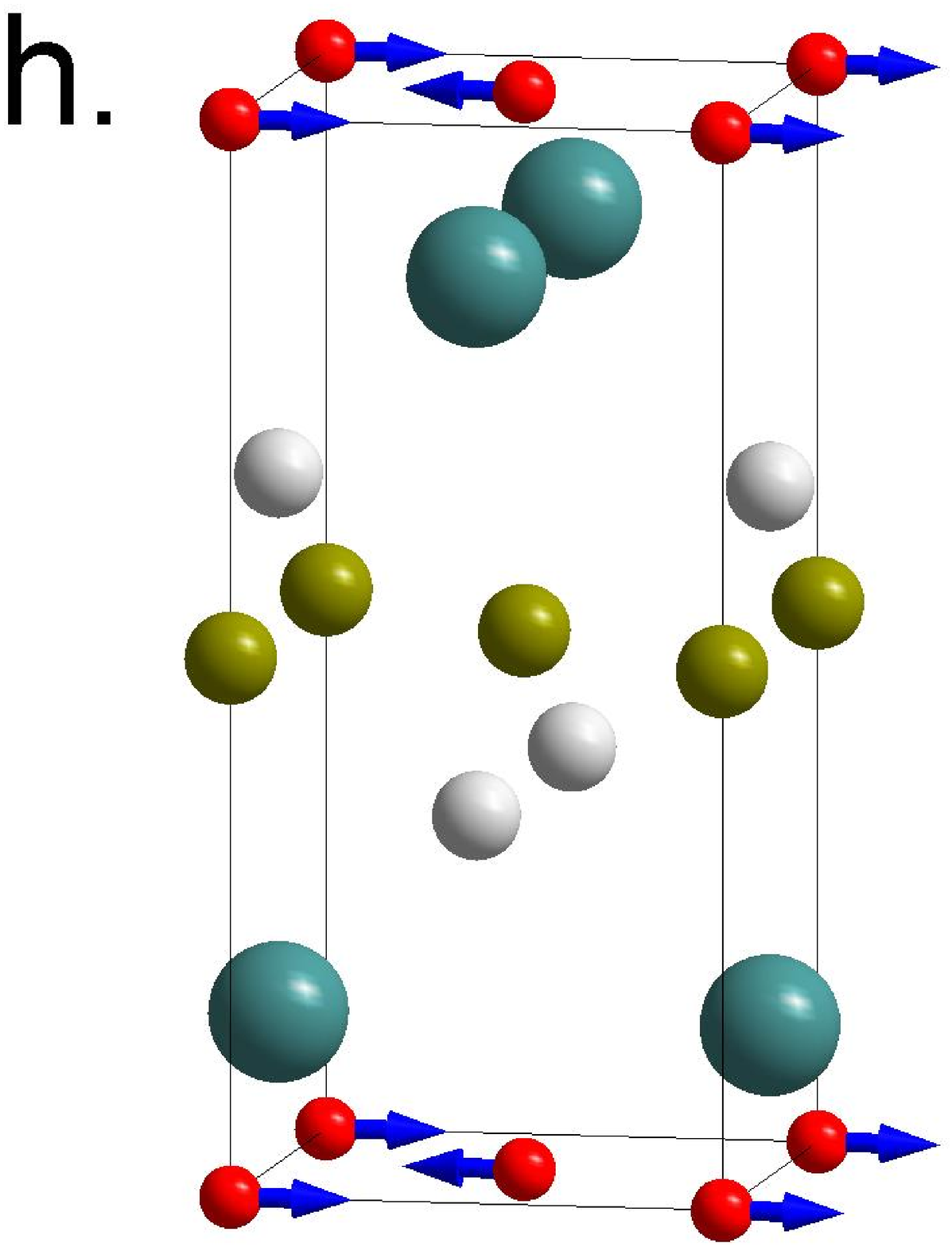}

\caption{(color online) Eight Raman-active modes of LaOFeAs:
(a)E$_g$ of La (96 cm$^{-1}$); (b)E$_g$ of As \& Fe (137
cm$^{-1}$); (c)A$_{1g}$ of La (161 cm$^{-1}$); (d)A$_{1g}$ of As
(not observed); (e)B$_{1g}$ of Fe (214 cm$^{-1}$); (f)E$_g$ of Fe
(278 cm$^{-1}$); (g)B$_{1g}$ of O (not observed); (h)E$_g$ of O
(423 cm$^{-1}$).} \label{fig3}
\end{figure}

\end{document}